\documentclass[aps,prl,showpacs,amssymb,nofootinbib,nobibnotes,10pt,superscriptaddress,twocolumn,longbibliography ]{revtex4-2}
\usepackage{tikz}
\usetikzlibrary{shapes,snakes,backgrounds,fit,decorations.pathreplacing}
\usepackage{bbm,times}
\usepackage{bm}
\usepackage{amsbsy}
\usepackage{amsthm}
\usepackage{amssymb}
\usepackage{amsfonts}
\usepackage{amsmath}
\usepackage[c]{esvect} 
\usepackage{dsfont} 
\usepackage{graphicx}
\usepackage[percent]{overpic}
\usepackage{subcaption}
\usepackage{epsfig}
\usepackage{epstopdf}
\usepackage{dsfont}
\usepackage{multibib}
\usepackage{color}
\usepackage{enumerate}
\usepackage{multirow}

\usepackage[colorlinks]{hyperref}
\makeatletter
\newcommand\org@hypertarget{}
\let\org@hypertarget\hypertarget
\renewcommand\hypertarget[2]{%
	\Hy@raisedlink{\org@hypertarget{#1}{}}#2%
}
\makeatother
\usepackage[figure,table]{hypcap}
\usepackage{MnSymbol}
\usepackage{enumerate}
\usepackage{float}
\hypersetup{
	bookmarksnumbered,   
	pdfstartview={FitH},
	citecolor={darkblue},
	linkcolor={darkred},
	urlcolor={darkblue},
	pdfpagemode={UseOutlines}}
\definecolor{darkgreen}{RGB}{50,190,50}
\definecolor{darkblue}{RGB}{0,0,190}
\definecolor{darkred}{RGB}{238,0,0}
\usepackage{soul}
\usepackage{CJKutf8}

\newcommand*\xoverline[2][0.75]{%
    \sbox{\myboxA}{$\m@th#2$}%
    \setbox\myboxB\null
    \ht\myboxB=\ht\myboxA%
    \dp\myboxB=\dp\myboxA%
    \wd\myboxB=#1\wd\myboxA
    \sbox\myboxB{$\m@th\overline{\copy\myboxB}$}
    \setlength\mylenA{\the\wd\myboxA}
    \addtolength\mylenA{-\the\wd\myboxB}%
    \ifdim\wd\myboxB<\wd\myboxA%
       \rlap{\hskip 0.5\mylenA\usebox\myboxB}{\usebox\myboxA}%
    \else
        \hskip -0.5\mylenA\rlap{\usebox\myboxA}{\hskip 0.5\mylenA\usebox\myboxB}%
    \fi}
\makeatother

\makeatletter
\renewcommand{\p@subsection}{}
\renewcommand{\p@subsubsection}{}
\makeatother

\usepackage{braket}

\usepackage{url}

\begin{document}

\title{Deterministic preparation of entangled Dicke states}

\author{Hui Wang }
\email{huiwangph@gmail.com}
\affiliation{Institute for Quantum Science and Engineering, Texas A\&M University, College Station, Texas 77843, USA}

\author{Marlan O. Scully}
\affiliation{Institute for Quantum Science and Engineering, Texas A\&M University, College Station, Texas 77843, USA}

\author{Girish S. Agarwal }
\email{Girish.Agarwal@ag.tamu.edu}
\affiliation{Institute for Quantum Science and Engineering, Texas A\&M University, College Station, Texas 77843, USA}
\affiliation{Department of Biological and Agricultural Engineering, Texas A\&M University, College Station, Texas 77843, USA}
\begin{abstract}
Dicke states $|J=N/2,m\rangle$ of a collection of $N$ atoms were central to Dicke's theory of superradiance. Except for the fully polarized end states, they are entangled many-body states; in particular, the single-excitation Dicke state is the W state well known in quantum information science. However, deterministic preparation of Dicke states with a prescribed spin projection remains challenging, especially without relying on postselection or heralding. Here we present a detuning-programmed Hamiltonian protocol using atoms or qubits in a dispersive cavity subject to a coherent transverse drive. In the absence of the drive, the off-resonant cavity produces an effective collective-spin interaction. By combining this cavity-mediated interaction with the coherent drive, tuning the atom--drive detuning enables the protocol, in principle, to target any allowed Dicke state along the symmetric Dicke ladder. Starting from the transverse-drive ground state, adiabatic ground-state interpolation prepares the selected Dicke state by ramping down the drive strength while ramping up the cavity-mediated interaction. After preparation, tuning the cavity into resonance with the atoms or qubits provides a direct way to probe the collective-emission response of the prepared Dicke state. We discuss implementation with superconducting circuit QED and show that the prepared states, especially the central Dicke state with $m=0$, provide resources for quantum sensing with Heisenberg-limited scaling.
\end{abstract}

\maketitle

\emph{Introduction.--} The controlled preparation of collective entangled states is a central goal across quantum information, many-body physics, and precision measurement. Among these, Dicke states form a paradigmatic family of permutation-symmetric many-body states with a fixed number of excitations~\cite{Dicke1954coherence,stockton2003characterizing}. They occupy a particularly rich position at the intersection of collective light--matter physics, multipartite entanglement, and quantum-enhanced sensing: the same symmetry that underlies enhanced collective emission also gives rise to entangled many-body states away from the fully polarized limits~\cite{toth2007detection,lucke2014detecting}. This combination of cooperative emission behavior and many-body entanglement makes Dicke states valuable resources for quantum information processing and quantum metrology, while providing a natural setting for exploring the interplay among collective emission, quantum correlations, and entanglement~\cite{prevedel2009experimental,pezze2018quantum}. Dicke states have been prepared or proposed in a broad range of platforms, including photonic systems~\cite{kiesel2007experimental,prevedel2009experimental,wieczorek2009experimental}, linear-optical schemes~\cite{thiel2007generation,kang2026heralded}, trapped ions~\cite{roos2004control,haeffner2005scalable,hume2009preparation,toyoda2011generation,noguchi2012generation}, and cavity- and circuit-QED platforms~\cite{fink2009dressed,mlynek2012demonstrating}. The single-excitation Dicke state is the $W$ state, which has been demonstrated in trapped-ion and cavity-QED systems~\cite{roos2004control,haeffner2005scalable,fink2009dressed,mlynek2012demonstrating}, while single photon subradiant states can be prepared by excitation with single photons~\cite{scully2015single}. For small $N$, the connection to familiar entanglement classes is especially transparent: the $N=2$, $m=0$ Dicke state is a Bell state, whereas the nontrivial $N=3$, $m=-1/2$ Dicke state belong to the $W$-type rather than the GHZ class~\cite{dur2000three}.

These advances demonstrate substantial control over collective quantum states, while deterministic preparation of a prescribed Dicke state remains an important goal. Detection-based, postselected, and heralded schemes have provided powerful routes to Dicke-state generation~\cite{kiesel2007experimental,prevedel2009experimental,wieczorek2009experimental,thiel2007generation,kang2026heralded}. For example, Ref.~\cite{thiel2007generation} proposed generating symmetric Dicke states of remote matter qubits through multiphoton far-field detection from initially excited atoms. Such approaches can herald successful preparation of the desired state, but remain probabilistic at the level of individual experimental runs, since success is conditioned on the required photon-detection events. Deterministic Hamiltonian approaches provide an alternative. More recently, rapid adiabatic passage has been explored as a route to Dicke-state preparation, using a frequency chirp to traverse a sequence of avoided crossings between neighboring Dicke states~\cite{carrasco2024Dicke}. This motivates complementary ground-state-engineering strategies in which the target Dicke state is encoded directly in the final many-body Hamiltonian.

Dicke superradiance is one of the paradigmatic manifestations of cooperative quantum dynamics, yet its relation to many-body entanglement is subtle. Dicke's original theory introduced the symmetric collective ladder of atomic states and the enhanced transition rates between them ~\cite{Dicke1954coherence,gross1982superradiance,agarwal2013quantum}. The largest Dicke transition rates occur near the center of the $J=N/2$ ladder. For even $N$, the central state $|J=N/2,m=0\rangle$ is a multipartite entangled state with half of the atoms excited. However, ordinary superradiant decay does not prepare this state as a pure state. Agarwal's master-equation treatment shows that the ensemble-averaged state is instead a diagonal incoherent mixture over Dicke states~\cite{agarwal1970master}. Moreover, the presence of entangled Dicke-state components does not by itself imply that the mixture is entangled: for ideal collective decay from the fully inverted product state, recent analyses show that the unconditioned state admits a separable description throughout the decay~\cite{wolfe2014certifying,rosario2025unraveling,bassler2025}. Thus, a superradiant burst signals collective emission and quantum correlations~\cite{auyuanet2010quantum}, but is not by itself a definitive witness of multipartite Dicke-state entanglement.

We therefore propose a cavity-QED protocol in which selected Dicke states are prepared by detuning-programmed adiabatic ground-state interpolation. An off-resonant cavity mediates an effective collective $\hat J_z^2$ interaction, while a transverse coherent drive connects an initial product state to the many-body ground state. Because the Hamiltonian is fully collective, the evolution remains within the $J=N/2$ Dicke manifold. For even $N$, the balanced protocol terminates in the central Dicke state $|J=N/2,m=0\rangle$. More generally, the atom-drive detuning shifts the minimum of the effective quadratic spin potential, thereby encoding the target projection $m_0$ directly in the final Hamiltonian. This provides a deterministic route to selected Dicke states within cavity- and circuit-QED architectures~\cite{fink2009dressed,mlynek2012demonstrating,blais2021circuit}. In this way, the Dicke ladder is converted from a collective decay pathway into a programmable manifold of pure many-body states. The prepared state can subsequently be probed through its collective-emission response, while its entanglement is a property of the prepared pure Dicke state itself rather than of the unconditioned superradiant decay dynamics~\cite{alabbar2026initiation}.

As an application, we consider quantum-enhanced rotation sensing. Quantum sensors exploit nonclassical correlations to push measurement precision beyond the standard quantum limit and, ultimately, toward the Heisenberg limit, with applications spanning inertial sensing and navigation, precision timekeeping, gravimetry and geodesy, and tests of fundamental physics~\cite{pezze2018quantum,bongs2019taking,ludlow2015optical, safronova2018search}. Realizing this quantum advantage requires many-body states whose entanglement can be controllably prepared and converted into enhanced parameter sensitivity. Dicke states provide a particularly important class of such resources, possessing metrologically useful multipartite entanglement~\cite{apellaniz2015detecting}.Entangled many-body states have already enabled enhanced rotation-angle estimation, precision spectroscopy approaching the Heisenberg limit, and quantum lock-in detection~\cite{meyer2001experimental,leibfried2004toward,zhang2026lockin}. For the family of Dicke states prepared here, we show that the central state with $m=0$ provides the highest sensitivity to transverse rotations and exhibits Heisenberg-limited scaling, $\Delta\vartheta\sim 1/N$~\cite{carrasco2024Dicke,holland1993interferometric,krischek2011useful}.



\emph{Adiabatic preparation of Dicke states.--}
We consider an ensemble of $N$ identical two-level atoms collectively coupled to a single cavity mode. The collective spin operators are defined as
\begin{equation}
\hat J_z=\frac{1}{2}\sum_{j=1}^{N}\hat{\sigma}_z^{(j)},\qquad
\hat J_+=\sum_{j=1}^{N}\hat{\sigma}_+^{(j)},\qquad
\hat J_-=\sum_{j=1}^{N}\hat{\sigma}_-^{(j)} .
\end{equation}
In the interaction picture, the atom--cavity interaction Hamiltonian is
\begin{equation}
\hat{H}_I(t) = g\left(\hat J_+ \hat{a} e^{-i\delta_c t} + \hat J_- \hat{a}^\dagger e^{i\delta_c t}\right),
\label{eq:HI_Dicke}
\end{equation}
where $\hat a$ is the annihilation operator of the cavity mode, $g$ is the atom--cavity coupling strength, and $\delta_c=\omega_c-\omega_0$ is the detuning between the cavity frequency $\omega_c$ and the atomic transition frequency $\omega_0$. We work in the far-detuned, dispersive regime, $|\delta_c| \gg gN$, where real cavity excitation is strongly suppressed and the cavity acts only as a virtual mediator~\cite{agarwal1997atomic, agarwal2013quantum}.

Writing $\hat{V}=g\hat J_+\hat{a}$, Eq.~\eqref{eq:HI_Dicke} becomes $\hat{H}_I(t)= \hat{V} e^{-i\delta_c t}+\hat{V}^\dagger e^{i\delta_c t}$. The first-order term averages to zero on timescales long compared with $1/|\delta_c|$. The leading nonvanishing contribution is therefore second order in $g/\delta_c$, and can be obtained by a Magnus expansion:
\begin{equation}
\hat{H}_{\rm eff}^{(2)}
=\frac{1}{\delta_c}[\hat{V}^\dagger,\hat{V}]
=\frac{g^2}{\delta_c}
\left(\hat J_-\hat J_+\hat{a}^\dagger \hat{a}-\hat J_+\hat J_-\hat{a}\hat{a}^\dagger\right).
\end{equation}
Projecting onto the cavity vacuum then yields
\begin{equation}
\hat{H}_{\rm eff}
=\langle 0|\hat{H}_{\rm eff}^{(2)}|0\rangle
=-\frac{g^2}{\delta_c}\hat J_+\hat J_- .
\label{eq:Heff_cavity_initial}
\end{equation}
Using the fact that $\hat J_+\hat J_- = \hat J^2-\hat J_z^2+\hat J_z$, and restricting to the symmetric Dicke manifold with $J=N/2$, where $\hat J^2$ has the eigenvalue $J\left(J+1\right)$,
Eq.~\eqref{eq:Heff_cavity_initial} can be written as
\begin{equation}
\hat{H}_{\rm eff}=\eta\left[\hat J_z^2-\frac{N}{2} \left(\frac{N}{2}+1\right)\right]-\eta \hat J_z ,
\label{eq:Heff_cavity_symmetric}
\end{equation}
with $\eta={g^2}/{\delta_c}$. For positive detuning, $\delta_c>0$, one has $\eta>0$, and the nonlinear term $\eta \hat J_z^2$ favors states with the smallest possible value of $|m|$.

We then apply a transverse classical drive with frequency $\omega_l$ and Rabi frequency $\Omega$. The detuning between the atomic transition and the drive is defined as $\Delta=\omega_0-\omega_l$. After transforming to a frame rotating at the drive frequency and making the rotating-wave approximation, the total Hamiltonian becomes
\begin{equation}
\hat{H}
=\eta\left[
\hat J_z^2-\frac{N}{2}\left(\frac{N}{2}+1\right)
\right]+(\Delta-\eta)\hat J_z+\Omega \hat J_x ,
\label{eq:Hrot_Dicke}
\end{equation}
where $\Omega$ denotes the transverse drive amplitude. For the protocol targeting the central Dicke state, which we refer to as the balanced protocol, we impose $\Delta(t)=\eta(t)$, so that the linear $\hat J_z$ term remains canceled throughout the interpolation. Up to an irrelevant additive constant, the working Hamiltonian is
\begin{equation}
\hat{H}(t)=
\Omega(t)\hat J_x+\eta(t) \hat J_z^2 .
\label{eq:Hwork_simple}
\end{equation}
For $\Omega>0$ and $\Omega\gg \eta N$, the transverse-field term dominates the Hamiltonian. The initial ground state is therefore the lowest eigenstate of $\hat J_x$, namely the fully symmetric spin-coherent
state polarized along the negative $x$-direction,
\begin{equation} 
|\Psi_0\rangle=\bigotimes_{j=1}^{N}|\phi_j\rangle ,
\qquad
|\phi_j\rangle =\frac{|e_j\rangle-|g_j\rangle}{\sqrt{2}} ,
\label{eq:initial_product_state} 
\end{equation}
where $\hat J_x|\Psi_0\rangle=-(N/2)|\Psi_0\rangle$. By contrast, when $\eta N\gg \Omega$, the nonlinear interaction dominates. For even $N$, the lowest-energy state of $\hat J_z^2$ in this manifold is the central Dicke state, $|D_N\rangle=|J=N/2,m=0\rangle$. 

Therefore, $|D_N\rangle$ can be prepared by starting from the product ground state of $\hat J_x$ in the transverse-field-dominated regime and then slowly decreasing $\Omega$ while increasing $\eta$, so that the ratio $\eta N/\Omega$ evolves from much smaller than unity to much larger than unity. Provided that the evolution occurs on a timescale long compared with the inverse many-body energy gap, the system adiabatically follows the instantaneous ground state and approaches $|D_N\rangle$. This state is the fully symmetric superposition of all configurations with $N/2$ excitations. After preparation, its collective-emission response can be probed by tuning the cavity into resonance and measuring the emitted field. 
For a Dicke state $|J,m\rangle$, the corresponding transition strength is proportional to $(J+m)(J-m+1)$~\cite{scully2015single}. For the central Dicke state $|D_N\rangle$, this reduces to $(N/2)(N/2+1)$.
For the numerical illustrations, we choose the linear ramps
\begin{equation}
\Omega(t)=\Omega_0\left(1-\frac{t}{T}\right),
\qquad
\eta(t)=\eta_0\frac{t}{T},
\end{equation}
with the detuning varied synchronously as $\Delta(t)=\eta(t)$ for the balanced protocol. Here $\Omega_0$ is the initial transverse-drive amplitude and $\eta_0$ is the final nonlinear-interaction strength. The linear ramp is chosen for simplicity; other sufficiently slow ramp profiles connecting the same initial and final Hamiltonians lead to the same adiabatic target state.
\begin{figure}[H]
\centering
\begin{minipage}[t]{0.49\columnwidth}
\centering
\begin{overpic}[width=\linewidth]{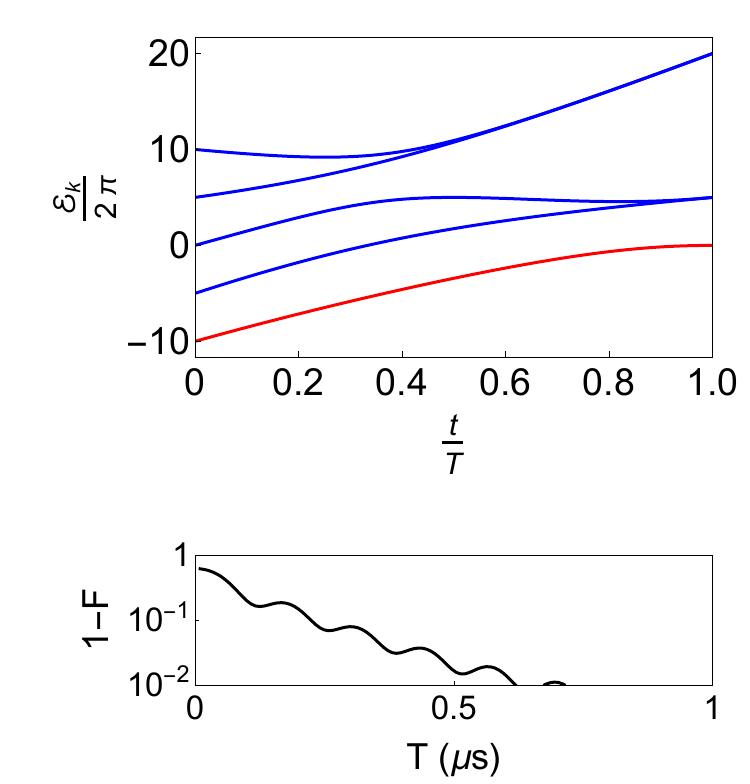}
\put(20,100){\scriptsize\bfseries (a)}
\end{overpic}
\end{minipage}
\hfill
\begin{minipage}[t]{0.49\columnwidth}
\centering
\begin{overpic}[width=\linewidth]{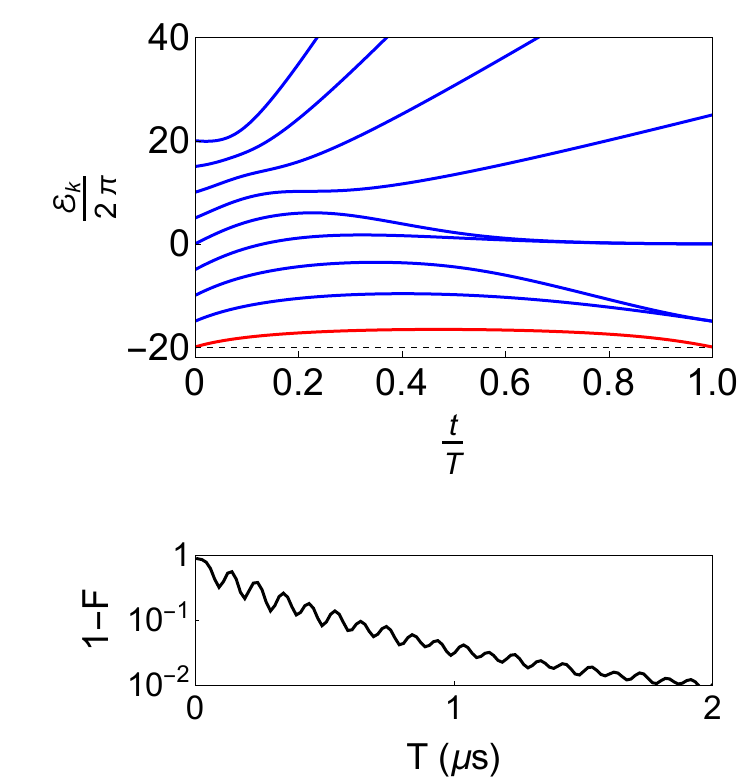}
\put(20,100){\scriptsize\bfseries (b)}
\end{overpic}
\end{minipage}
\caption{
Instantaneous spectra and final infidelities for adiabatic Dicke-state preparation. In the upper panels, the vertical axis shows $\mathcal{E}_k/(2\pi)$, where $\mathcal{E}_k$ is the $k$th instantaneous eigenenergy of the ramp Hamiltonian, ordered from lowest to highest energy. Frequencies are given in $\mathrm{MHz}$. The red curve denotes the ground-state branch, while the blue curves denote excited-state branches. The lower panels show the final infidelity $1-F$ as a function of the total ramp time $T$. 
(a) Dicke-state preparation with target $|J=2,m=0\rangle$. 
(b) Dicke-state preparation with target $|J=4,m=-2\rangle$.
}
\label{fig:spectrum}
\end{figure}

For the numerical example, we take
$\Omega_0=\eta_0=2\pi\times 5~{\rm MHz}$ and $J=2$, corresponding to $N=4$ atoms. As shown in Fig.~\ref{fig:spectrum}(a), the instantaneous ground-state branch remains separated from the excited-state branches throughout the ramp, providing a finite gap for adiabatic following.

To verify that the state dynamically follows this adiabatic path, we also solve the time-dependent Schr\"odinger equation for different total ramp times $T$, starting from the initial ground state $|\Psi_0\rangle$. The instantaneous spectra show that the ground-state branch continuously connects the initial spin-coherent state to the target Dicke state. However, the physical state reaches this target only when the ramp is sufficiently slow to suppress nonadiabatic transitions to the excited branches. This mechanism is distinct from the rapid-adiabatic-passage protocol of Ref.~\cite{carrasco2024Dicke}, where an independent frequency chirp traverses a sequence of avoided crossings between neighboring Dicke states to reach the desired Dicke state. Here, instead, the target is encoded as the minimum of the final quadratic spin potential and is reached by continuously following the instantaneous ground-state branch. The lower panel of Fig.~\ref{fig:spectrum}(a) shows the final infidelity $1-F$, where
\begin{equation}
F=|\langle J=2,m=0|\psi(T)\rangle|^2 .
\end{equation}
As $T$ is increased, the final infidelity decreases overall, confirming that the system approaches the target state in the adiabatic limit. The small oscillations reflect finite-time nonadiabatic transitions during the ramp.

\emph{Preparation of arbitrary Dicke states.--}
The above scheme can be generalized within the same Dicke manifold to prepare a target state $|J,m_0\rangle$. A key ingredient in this generalization is the term linear in $\hat J_z$ in Eq.~\eqref{eq:Hrot_Dicke}. Instead of imposing $\Delta(t)=\eta(t)$, we retain this term and tune its coefficient through the atom–drive detuning, thereby selecting the target Dicke projection.
\begin{equation}
\hat H(t)=\Omega(t)\hat J_x
+\eta(t)\left[\hat J_z^2-J(J+1)\right]
+\eta(t)(\beta-1)\hat J_z .
\label{eq:Hwork_beta}
\end{equation}

The role of the detuning is most transparent at the end of the ramp, where $\Omega(T)=0$ and $\eta(T)=\eta_0$. Since the Dicke states $|J,m\rangle$ are eigenstates of $\hat J_z$, their endpoint energies are
\begin{equation}
\frac{E_m(T)}{\eta_0}
=
m^2-J(J+1)+(\beta-1)m .
\end{equation}
To select a target state $|J,m_0\rangle$, we choose $\beta=1-2m_0$. Accordingly, the detuning is varied throughout the ramp as
\begin{equation}
\Delta(t)=\eta(t)(1-2m_0),
\end{equation}
thereby keeping the detuning bias matched to the chosen target throughout the interpolation. With this choice, the endpoint energies become
\begin{equation}
\frac{E_m(T)}{\eta_0}
=
(m-m_0)^2-m_0^2-J(J+1),
\end{equation}
which is minimized at $m=m_0$. Thus, the target Dicke state is selected by the detuning through the position of the minimum of the final quadratic spin potential. Equivalently, for a target state with $n$ excited atoms, $m_0=n-N/2$, so that $\beta=N+1-2n$.

Figure~\ref{fig:spectrum}(b) illustrates the detuning-selected protocol for $J=4$ and $m_0=-2$, using the same
$\Omega_0=\eta_0=2\pi\times5~{\rm MHz}$. In this case $\beta=5$, and the final spin potential is minimized at
$m=-2$. The red curve shows the instantaneous ground-state branch selected by this detuning, while the lower panel shows that the final infidelity decreases overall as the ramp time $T$ increases, confirming adiabatic following toward $|J=4,m=-2\rangle$. For the parameters used here, both protocols approach their target Dicke states on microsecond timescales, with the detuning-selected case exhibiting slower convergence. Additive energy constants have been omitted in the plotted spectra. Consequently, the near coincidence of the initial and final ground-state energies in Fig.~\ref{fig:spectrum}(b) does not imply that the corresponding states are similar; the change in the state itself is shown more directly by the Husimi $Q$-function below.


Collective cavity-QED interactions have long been explored for the generation of multipartite entangled states~\cite{zheng2001one}, while superconducting circuits provide a versatile platform for realizing atomic-physics and quantum-optics phenomena on chip~\cite{you2011atomic}. Circuit QED is especially well suited to the present protocol, since collective spin interactions and transverse-drive control of the required form have already been realized in multiqubit devices. Resonator-mediated one-axis-twisting dynamics have been implemented with up to 20 superconducting qubits coupled through a common bus resonator, demonstrating collective nonlinear interactions in a large multiqubit device and enabling the preparation of a GHZ state formed by a coherent superposition of the collective ground and fully excited states~\cite{song2019generation}. More recently, six Xmon qubits coupled through virtual-photon exchange mediated by a common resonator realized an effective Hamiltonian containing both a transverse collective field and a quadratic $\hat J_z^2$ interaction~\cite{zheng2025experimental}. These results establish access to MHz-scale collective interactions $\eta$ together with fast transverse-drive control in multiqubit circuit-QED devices.

Fast control ramps on the order of $10^2~\mathrm{ns}$ have been demonstrated in closely related multiqubit circuit-QED experiments. More generally, strategies for suppressing dephasing in superconducting qubits, including capacitively shunted flux-qubit designs, have also been proposed~\cite{you2007low}. In the multiqubit experiments relevant here, dephasing times are typically a few microseconds, while energy-relaxation times are of order ten microseconds or longer~\cite{zheng2025experimental,zhang2022synthesizing}. Consequently, the state preparation can be completed on a timescale much shorter than these decoherence times, thereby effectively mitigating decoherence during adiabatic evolution. Thus, decoherence is not expected to preclude the preparation of the Dicke states proposed here, and the illustrative parameters in Fig.~\ref{fig:spectrum} lie within an experimentally accessible regime.



The instantaneous state evolution can be visualized using the Husimi $Q$-function,
\begin{equation}
Q_t(\theta,\phi)=|\langle \theta,\phi|\psi(t)\rangle|^2,
\end{equation}
where $|\psi(t)\rangle$ is the time-evolved state. At the beginning of the ramp, it is the spin-coherent state polarized along the negative $x$-direction, for which
\begin{equation}
Q_{t=0}(\theta,\phi)
=
\left(
\frac{1-\sin\theta\cos\phi}{2}
\right)^{2J}.
\end{equation}

\begin{figure}[h]
    \includegraphics[width=\linewidth]{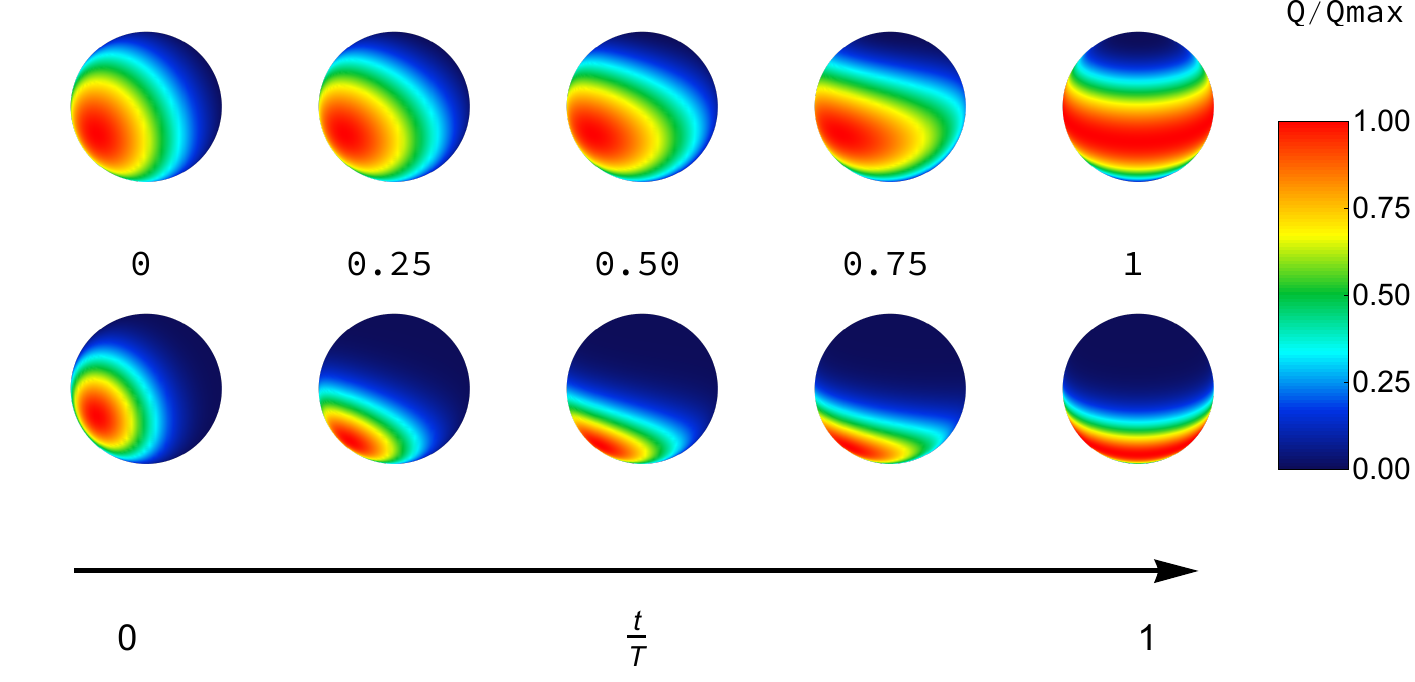}
    \caption{Time evolution of the Husimi $Q$-function during adiabatic preparation for $t/T=0,0.25,0.50,0.75,1$. The upper and lower rows
    correspond to targets $|J=2,m=0\rangle$ and $|J=4,m=-2\rangle$, respectively. Each sphere is normalized by its own maximum $Q_{\max}$. A total ramp time of order $T\sim1~\mu{\rm s}$ is representative of the preparation timescales as shown in Fig.~\ref{fig:spectrum}.
    }
    \label{fig:qfunction}
\end{figure}

As $\eta(t)/\Omega(t)$ increases, the distribution evolves toward the Dicke state selected by the final spin potential. At the end of the ramp,
\begin{equation}
Q_{t=T}^{(m_0)}(\theta,\phi)
\propto
\left(\cos\frac{\theta}{2}\right)^{2(J+m_0)}
\left(\sin\frac{\theta}{2}\right)^{2(J-m_0)} ,
\end{equation}
which is independent of $\phi$ and peaked near $\cos\theta=m_0/J$. Figure~\ref{fig:qfunction} shows this evolution for the two protocols of Fig.~\ref{fig:spectrum}. For the balanced protocol, the distribution evolves toward the characteristic equatorial structure of $|J=2,m=0\rangle$, whereas for the detuning-selected protocol it approaches the latitude corresponding to $|J=4,m=-2\rangle$, with $\cos\theta=-1/2$. Figure~\ref{fig:qfunction} thus provides a phase-space visualization of how the instantaneous state evolves toward the Dicke projection selected by the detuning.

\emph{Rotation sensing with prepared Dicke states.--} 
We now quantify the transverse rotation sensitivity of the prepared Dicke state $|J,m_0\rangle$. The selected value of $m_0$ determines its transverse collective-spin fluctuations and hence the quantum Fisher information for rotations about axes perpendicular to the quantization axis.

After preparation, we use the prepared Dicke state as a
probe for local estimation of a rotation angle $\vartheta$ generated by a collective spin component in the transverse plane,
\begin{equation}
|\psi_\vartheta\rangle=e^{-i\vartheta\hat G}|J,m_0\rangle,
\qquad
\hat G=\mathbf n\cdot\hat{\mathbf J},
\end{equation}
where $\mathbf n=(\cos\alpha,\sin\alpha,0)$ specifies the rotation axis in the plane transverse to the quantization axis. Here the rotation is unitary and the probe state remains pure. The quantum Fisher information is therefore $F_Q=4(\Delta \hat G)^2$, where $(\Delta \hat G)^2=\langle \hat G^2\rangle-\langle \hat G\rangle^2$~\cite{braunstein1994statistical,pezze2018quantum,agarwal2022quantifying}. The corresponding quantum Cram\'er--Rao bound gives the quantum-limited sensitivity
$\Delta\vartheta_{\rm Q}\ge 1/\sqrt{\nu F_Q}$, where $\nu$ is the number of independent repetitions.

For the Dicke state $|J,m_0\rangle$, one has $\langle \hat G\rangle=0$. Owing to its rotational symmetry about the quantization axis, the transverse fluctuations are independent of the azimuthal orientation $\alpha$, with $\langle \hat G^2\rangle=[J(J+1)-m_0^2]/2$. The quantum Fisher information therefore becomes
\begin{equation}
F_Q(m_0)
=2\left[J(J+1)-m_0^2\right]
=\frac{N(N+2)}{2}-2m_0^2 .
\label{fq}
\end{equation}
Equation~\eqref{fq} is maximized at $m_0=0$, corresponding to the central Dicke state. For this state, $F_Q=N(N+2)/2$, and the quantum Cram\'er--Rao bound gives
\begin{equation}
\Delta\vartheta_{\rm Q}
\ge\sqrt{\frac{2}{\nu N(N+2)}} \simeq
\frac{\sqrt{2}}{\sqrt{\nu}\,N},
\end{equation}
where the last expression holds for large $N$. For a single experimental repetition, $\nu=1$, this gives $\Delta\vartheta_Q\simeq\sqrt{2}/N$. This corresponds to sub-shot-noise rotation sensitivity, improving over the standard quantum limit by a factor $\sqrt{N/2}$, while attaining the Heisenberg-limited scaling set by collective-spin metrology. The central Dicke state prepared by the balanced protocol is therefore the optimal member of this family for transverse rotation sensing.



\emph{Conclusion.--}
We have proposed a deterministic cavity-QED protocol for preparing selected Dicke states by detuning-programmed adiabatic ground-state interpolation. An off-resonant cavity generates an effective collective interaction, while a transverse coherent drive connects an initial spin-coherent product state to the target many-body ground state. For even $N$, the balanced protocol prepares the central Dicke state $|J=N/2,m=0\rangle$, whereas tuning the detuning shifts the minimum of the effective quadratic spin potential and thereby selects states with different values of $m_0$. In contrast to unconditioned superradiant decay, the protocol prepares a selected pure Dicke state whose collective-emission response can subsequently be probed by tuning the cavity into resonance. The required collective interactions and coherent control are compatible with capabilities already demonstrated in multiqubit circuit-QED platforms. The prepared states also provide metrological resources for transverse rotation sensing, with the central $m=0$ state exhibiting Heisenberg-limited scaling. More broadly, the ability to program a chosen Dicke state and subsequently access its collective-emission and metrological response within the same architecture provides a route toward integrating many-body state engineering, cooperative quantum optics, and quantum-enhanced sensing.

G.S.A. thanks Prof. Dawei Wang and Dr. C. Song for discussions during the initial stages of this work. This work was supported by the Robert A. Welch Foundation (Grant Nos. A-1943-20240404 and A-1261) and the Department of Energy, Fusion Energy Science Program (Award No. DE-SC0024882: IFE-STAR).

\bibliography{Dicke.bib}

\end{document}